\documentstyle[twocolumn,aps]{revtex}

\begin{document}
\draft
\title{  BDSW protocol revisited: an efficient method for the key 
distillation without classical computational complexity}
\author{Xiang-Bin Wang\thanks{Email address: wang@qci.jst.go.jp}\\
IMAI Quantum Computation and Information Project,
ERATO, JST,
Daini Hongo White Bldg. 201, \\5-28-3, Hongo, Bunkyo,
Tokyo 133-0033, Japan}
\maketitle
\begin{abstract} 
In quantum key distribution(QKD), one can use
a classical CSS code to distill the final key.
However, there is a constraint for the two codes in CSS code
and so far it is unknown
how to construct a large CSS code efficiently. 
Here we show that the BDSW method given by Bennett et al can be modified 
and  the error correction and privacy amplification can be done
separately with two {\it independent}  parity matrices.
With such a modification, BDSW method can be used to distill the final key
without any classical computational complexity. 
We also apply the method to the case of  imperfect source where
a small fraction of signals are tagged by Eve.
\end{abstract}
{\it Introduction.}
Quantum key distribution (QKD)
\cite{bene,eker,BDSW,ben2,ben3,brus,gisi}
could be the one that is closest to immediate application in practice
because of its relative low technical overhead:
the only thing required there is preparation, transmission and measurement
of a 2-level quantum state, e.g., a single photon.  
\\
The security proof of QKD  is strongly non-trivial. 
However, it is greatly simplified if we look at the problem from the entanglement 
distillation viewpoint. The first protocol for the entanglement distillation was 
given by Bennett et al\cite{BDSW,ben2}, BDSW protocol. It was then pointed out by
Deutsch et al\cite{ekert} that the distillation protocol can be used for
secure QKD: we can first purify the entangled pairs and then take measurement
in the same basis on each side.  Latter on,
it was shown that\cite{lo3,squeeze,gl} actually the $fidelity$ result of distillation protocol\cite{BDSW} 
is $always$ correct given whatever initial state of the raw pairs: Suppose in the case that each raw 
pairs are in a Bell state, the 
protocol may distill out $m$ pairs in a state $\rho$ whose fidelity to $m$
perfect entangled pairs is almost 1. As it has been shown by Lo and Chau\cite{lo3}, 
if the fidelity is exponentially close to 1, then Eve's information
is exponentially small.
In the most general case, we imagine a Bell measurement on each pair
just before the distillation then we obtain the same $\rho$ after distillation.
In the distillation, two remote parties,
Alice and Bob need the local controlled-NOT gate on each side
to collect the parity information of a random subset of the raw pairs
into one pair (destination pair) and then measure the parity 
of that 
pair and discard the destination pair. Note that the parity measurement is
a collective measurement of $M_iM_i$.
They repeat this step until they they
can compute the location of all flipping errors of the remained $m$ pairs.
The initial Bell measurement commutes with all operations in distillation 
therefore can 
be postponed until the end of the distillation. Moreover, after this
delay we can even remove the step of Bell measurement: 
with this removal, they finally
obtain $m$ pairs in a different state, $\rho'$, 
but $\rho$ and $\rho'$ have the same fidelity 
to $m$ perfect entangled pairs. And the fidelity value is the only thing we
are caring about here. 
Moreover, since all destination pairs have been discarded, it 
does not affect the fidelity of $\rho'$ if
they then take local measurement $M_i\otimes M_i$ on each side
to those discarded pairs and then announce the outcome. 
Since this local measurement commute with the parity measurement
$M_iM_i$, they can exchange the 
order of them therefore  measurement
$M_iM_i$
is unnecessary: once they announced the specific result of local measurement,
they have known the parity already. Therefore all they need there is just
local operation and classical communication (LOCC), this is just BDSW protocol.
\\
Therefore, in doing the entanglement purification or privacy amplification,
we can safely assume that each raw pairs are in one of the 4 Bell states,
$|\phi^\pm\rangle=\frac{1}{\sqrt 2}(|00\rangle\pm|11\rangle),
|\psi^\pm \rangle=\frac{1}{\sqrt 2}(|01\rangle\pm|10\rangle)$. 
Or equivalently,
we can assume Pauli channel for the qubit transmission:
\begin{eqnarray}
\sigma_x= ( \begin{array}{cc} 0 & 1  \\
                                       1 & 0
             \end{array} ),
 \sigma_y= ( \begin{array}{cc} 0 & -i  \\
                                            i &  0
             \end{array} ),
\sigma_z= ( \begin{array}{cc} 1 & 0  \\
                                         0 & -1
             \end{array} )\end{eqnarray}
If Alice starts from $N$ pairs of $|\phi^+\rangle$ state, 
the bit-flip rate is defined as the percentage of pairs
which have been changed into state 
$|\psi^+\rangle$ or state 
$|\psi^-\rangle$; 
phase-flip rate is defined as the percentage of pairs
which have been changed into state 
$|\phi^-\rangle$ or state 
$|\psi^-\rangle$. Equivalently, 
given Pauli channel,
    the channel operation $\sigma_x$ or $\sigma_y$ causes a bit-flip,
the channel operation $\sigma_y$ or $\sigma_z$ will cause a phase-flip.

{\it The constraint in classical CSS code.}
The entanglement distillation can also be done by a type of
quantum error correction code: the CSS code which is named by it's inventors,
Calderbank, Shor and Steane\cite{css1,css2}. They use classical linear codes
$C_1,C_2^\perp$ to correct bit-flip errors and amply the privacy of the 
final key\cite{shorpre}. Here the error correction (EC) and privacy amplication (PA)
are decoupled because of the constraint: 
\begin{eqnarray}\label{cons}C_2\subset C_1.
\end{eqnarray} 
Due to this constraint, it is so far not known on how to construct
a large classical CSS code efficiently. 
  Especially, the construction
task is even more complicated when we have
another constraint: $C_1$ must be efficiently decodable, e.g., Spielman
code\cite{decodable}.\\
It is not a problem to construct small CSS codes and we can distill the final
key concatenatedly. However, this method seems not economic:
when we use small CSS codes, we have to assume a large statistical fluctuation
therefore 
decrease the key rate. Another choice is to use two-way classical
communication in key distillation\cite{gl}, however, the key rate is
also low.
\\
If we use BDSW protocol\cite{BDSW} with one-way random hashing, 
we do not need CSS code, we only need
a random matrix, or a linear code, $C$.  
However, in the present form of BDSW protocol\cite{BDSW}, 
the error correction and 
privacy amplification are combined together. Whenever any hashing step
is done, there are backward actions therefore the remained pairs 
cannot be simply described by the criteria of ``flipping rate''. 
To such a case,
the efficiently decodable error correction code given by 
Spielman\cite{decodable} cannot be directly applied. 
If we use Shanon code for BDSW protocol, the decoding is
 complicated.
\\
{\it In short, there are computational difficulties in
both BDSW protocol\cite{BDSW} and CSS code\cite{shorpre}
for key distillation 
in practice: the $construction$ difficulty
 in using CSS code and the 
decoding complexity in using BDSW protocol.}
In this paper, we modify BDSW method so that the  
EC and PA are treated separately with two $independent$ linear codes.
 Moreover, after the modification,
we can choose to use Spielman's code\cite{decodable} for EC therefore
 error correction step is done efficiently. 
There is no computational complexity in PA
step of our protocol since we don't need to really correct the phase-flip
errors.
 
{\it Modified BDSW protocol: error correction
and privacy amplification with two {\bf independent} parity matrices.}
 We now consider
to modify BDSW protocol\cite{BDSW} therefore EC and PA can be done separately
with two independent parity matrices.
In a previous work given by Lo\cite{ld}, the BDSW
method is modified so that the error
correction and privacy amplification can be done separately with two
independent matrices.
However, there, a pre-shared secret string used as one-time-pad is requested. 
Here we don't use one-time-pad. 
Our modified protocol here is laso different from
the hashing method given by Ref\cite{lo33} where the constraint
of eq.(\ref{cons}) is used and the construction difficulty still exists.
\\
Given $N$ raw pairs, we can use two $N-$bit binary strings, the bit string
$s_{b}$ and the phase string $s_{p}$ 
to represent the quantum state of these raw pairs. Given any raw pair,
if it bears a bit-flip, the corresponding element 
in string $s_{b}$ is 1, otherwise it is 0; if it bears
 a phase-flip, the corresponding element  in string $s_{p}$ is 1, otherwise
it is 0.
For example, if the raw pairs are in the state 
$|\phi^+\rangle|\phi^+\rangle|\psi^+\rangle|\phi^-\rangle
|\psi^-\rangle$, the two classical strings are
\begin{eqnarray}
s_{b}= 00101;s_p=00011.
\end{eqnarray}
One can see that, the state of raw pairs is explicitly known if both bit 
string
and phase string are known. In the BDSW protocol\cite{BDSW} for
entanglement distillation, 
there are many steps of parity measurement, after each step, 
we discard one pair
therefore we have a new shortened 
strings for the remained pairs. Our purpose is to know exactly the strings for
the remained pairs after some hashing steps.
For clarity, we shall use $s_{bi},s_{pi}$ to represent the strings of the remained
pairs after step $i$,
e.g., we use $s_{b0},s_{p0}$ for the initial strings.
\\
Suppose the bit-flip
rate is less than $\delta_b$ and the phase-flip rate is less than
$\delta_p$. Initially the number of likely string for $s_{b0}$ and $s_{p0}$ 
are
less than $w_b$ and $w_p$, respectively. Here $w_b,w_p$ are determined by
$\delta_b,\delta_p$:
\begin{eqnarray}
\omega_{b}=2^{N\cdot H(\delta_b)};\omega_{p}=2^{N\cdot H(\delta_p)}
\end{eqnarray}
 and
\begin{eqnarray}
H(x)=-x\log_2 x -(1-x)\log_2 (1-x).
\end{eqnarray}
After we check the parities in $Z$ basis with an efficiently 
decodable linear code $C$\cite{decodable} 
which corrects  $\delta_b N$ errors, we can compute the
locations of all bit-flip errors and flip them back. Note that after
this error-correction, string $s_p$ changes due to the backward effect.
However, the number of likely string for $s_p$ is still upper bounded by
$\omega_p$, the number of likely string for $s_{p0}$. To do phase error correction
is difficult because of the backward effect. But to do privacy amplification
is simple: Once we know the number of likely
string for $s_p$, we {\it in principle}
know how to correct all phase errors and this in principle computability is enough.          
\\ There are two independent 
elementary operations in the modified BDSW protocol:
\\(1) Error correction: At step $i$, the classical binary 
strings for the remained
 $N-i$ qubits are $s_{bi},s_{pi}$ with $i$ being started from 0. Alice and Bob
generate a random binary string $r_i$ and measure the parity value
 of $r_i\cdot s_{bi}$ at each side and announce the results. The can do the 
measurement by first collect the parity of all pairs indicated by non-zero
elements in $r_i$ to the destination pair, $d_i$ and then measure pair
 $d_i$ in $Z$ basis at each side. In collecting the parity into pair
$d_i$, they only
need to do the controlled-NOT operations at each side (bi-CNOT) with pair
$d_i$ being the target pair and all other pairs indicated by $r_i$ being
the controlled pair. They then discard pair $d_i$. 
\\
If the initial bit-flip error rate is $\delta_b$, the number of
likely strings for $s_{b0}$ is $2^{NH(\delta_b)}$. They need run the step for
$n_b=NH(\delta_b)$ times to compute the explicit form the string
$s_{bn_b}$ and then Bob takes bit-flip operation to those
qubits bearing a bit-flip error. The process can be summarized
by error correction through an $n_b\times N$ random matrix.
Decoding such a random matrix could be very complicated.
However, {\it we can first randomly permute the qubits
and then replace the random matrix by Spielman code \cite{decodable}
which can be decoded efficiently.}
\\
We now consider the backward action. 
We denote $|\phi^{+}\rangle,
|\phi^{-}\rangle, 
|\psi^{+}\rangle,
|\psi^{-}\rangle$ by $|\chi_{00}\rangle,
|\chi_{01}\rangle,|\chi_{10}\rangle, |\chi_{11}\rangle$, respectively.
Given two pair state $|\chi_{a,b}\rangle|\chi_{a',b'}\rangle$, if we do bi-CNOT on this two 
pairs with the second pair being the target, we have
\begin{eqnarray}
|\chi_{a,b}\rangle|\chi_{a',b'}\rangle
\longrightarrow |\chi_{a,b\oplus b'}\rangle|\chi_{a'\oplus a,b'}\rangle.
\end{eqnarray}
Here $\oplus$ is bitwise sum and each of $a,b,a',b'$ can only take 0 or 1.
This shows that, given
$s_{pi}$, string $s_{pi+1}$ is determined exactly since the backward
action is only determined by phase flip information of the destination pair,
pair $d_i$. That is to say, the so-called backward action does not change
the number of likely string of $s_{pi}$. If the initial phase-flip rate
is $\delta_p$, then the number of likely string for each $s_{pi}$ is fixed
at $\omega_{p0}$. After doing error correction, they can then
start the second elementary step:
\\(2) Privacy amplification. If they can also locate all positions of
phase-flip errors, they can then flip them back and obtain pure entangled
pairs therefore complete the entanglement distillation.  
At step $j$, the classical binary 
strings for the remained
 $N-j$ qubits are $s_{bj},s_{pj}$ with $j$ being started from $n_b$. 
Alice and Bob
generate a random binary string $r_j$ and measure parity value
 of $r_j\cdot s_{pj}$ at each side and announce the results. The can do the 
measurement by first collect the  parity in $X-$basis of all pairs 
indicated by 
non-zero
elements in $r_j$ to the destination pair, $d_j$ and then measure pair
 $d_j$ in $X$ basis in each side. To collect the parity, they only
need to take bi-CNOTs in $X-$basis with pair
$d_j$ being the target pair and all other pairs indicated by $r_j$ being
the controlled pair. They then discard pair $d_j$. And they use  
 binary string $s_{bj+1},s_{pj+1}$ to represent the remained $N-j-1$ qubits.
It has been proven\cite{BDSW} that, they only need repeat the step for
\begin{eqnarray}
n_p=N\cdot H(\delta_p)
\end{eqnarray}
times in order to specify the final string of $s_p$. 
However, since their only purpose is to obtain a secure final key, they need not
take phase-flips to those pairs bearing a phase-flip error. Instead, they 
may directly measure the remained pairs after $n_p$ steps of parity measurement in
$X$-basis.
Moreover, since all destination pairs are discarded, the measurements in
$X$ basis to them are also unnecessary. 
The only thing now remained here is the 
bi-CNOTs in $X-$basis. 
These are equivalent to bi-CNOTs in $Z-$basis with the target 
pair and the controlled pairs being reversed. 
Therefore, all operations needed in the distillation
are done in $Z$ basis
and Alice can replace the initial distribution of entangled pairs by
sending Bob single qubits randomly chosen from the BB84 set 
$\{|0\rangle, |1\rangle,|\pm\rangle=\frac{1}{\sqrt 2}(|0\rangle\pm|1\rangle)\}$.
In each step $j$, they simply replace each bits in the set indicated by $r_j$
by the parity of that bit and bit $d_j$ and discard the bit $d_j$.
After error correction and privacy amplification, the  remained bits
can be used as the final key and the key rate is
\begin{eqnarray}
R=1-H(\delta_b)-H(\delta_p).
\end{eqnarray}
{\it QKD with imperfect source}. 
Having removed the computational complexity by the modifying 
BDSW method\cite{BDSW},
we now consider a type of physical imperfection of source. In practice,
it's very often to use the weak coherent states in stead of a real single 
photon source, which is a difficult technique. However, there will be
a small fraction of multi-photon signals if we use weak coherent states.
To those multi-photon signals, Eve may first split the light beam, 
keep one photon with her and send other photons to Bob. She will wait
until Alice announces the measurement basis of that signal. Such a photon
number splitting (PNS) attack will help Eve to have full information
of bit values of multi-photon signals without disturbing it at all. 
More generally, we can
use the ``tagging'' model\cite{gllp} to describe the type of
imperfect source: Alice uses perfect single photon source but she tells
Eve the exact states of a fraction of her signals . 
That is to say, Eve may tag a few of qubits 
without disturbing them at all. Here, we treat the issue in a similar way 
given by ref\cite{gllp}, but we shall not use CSS code therefore we don't 
have the $construction$ complexity in practice.
\\
For clarity, we consider the entanglement distillation first.
Initially, Alice prepares a number of
perfect entangled pairs, $|\phi^+\rangle$. 
Before entanglement distribution, a small fraction $\Delta$ of them
are tagged by Eve, i.e., Alice measures these tagged pairs in $Z$ basis or $X$
basis and tells Eve her measurement bases and outcome. 
And latter, in error test
or key distillation, those tagged pair will be , 
only measured it in the same basis used by Alice before 
entanglement distribution. 
If a tagged
pair was measured in $Z$ basis, the averaged phase-flip error is a half,
after passing through whatever noisy channel. The proof for this is very 
simple. According to its definition, $if$ the measure the pair
at each side in $X-$basis and obtain different outcome,
  then we say that pair bears a phase-flip error. Consider the
tagged state $|00\rangle$ or $|11\rangle$. $If$ Alice now measures her qubit
 in 
$X-$basis, the outcome is totally random and has no correlation with any 
other qubit. No channel can create correlation between a qubit
and a random result.
Therefore, if Bob measures his qubit in $X-$basis, he must have half a chance
to obtain a different outcome, since otherwise his qubit has non-zero 
correlation with Alice's qubit. \\Now let's consider the error test.
 The measured
error rate in $X-$basis on the test pairs does not indicate the correct value
of phase-flip rate of untested pairs. 
(Here we only consider the asymptotical result. We
don't consider the statistical
fluctuation, for simplicity.) 
Since Eve can treat tagged pairs and untagged pairs differently, we have to 
consider them separately. For those untagged pairs, since the measurement 
basis of each pair is unknown to Eve, the result bit errors
of the test pairs in $X-$basis on 
untagged pairs can be used to indicate the correct phase-flip error rate
of the remained untagged pairs. However, for those tagged pairs, the 
situation is different. Alice's measurement basis has been pre-determined
and announced,
they cannot choose basis randomly latter. The measurement outcome of
tagged pairs in $X-$basis does not indicate anything about the 
phase-flip rate of
those untested tagged pairs in $Z-$basis. 
In fact, the phase-flip rate of 
tagged pairs in $Z-$basis is fixed to $50\%$ while the bit-flip rate of 
those $X-$basis tagged
pairs can be 0, as Eve likes. This shows that, if the $tested$ bit-error rate
in $X-$basis 
is $\delta_p$, it could be the worst case that the phase-flip rate
of the untagged pairs is $\delta_p/(1-\Delta)$ and the phase-flip rate to
tagged pairs is fixed at $50\%$.
Differently, bit-flip error rate can be indicated 
correctly by the error test. Because in this case, the test pairs and the 
untest pairs will be measured in the same basis. Before they do the test,
Eve cannot tell which pairs will be used as the test pairs.
Suppose after all error tests, there are $N$ pairs remained. 
\\In bit-flip error correction,
 the backward action is 
different from that of the perfect source. Say, if the bit-flip rate is
$\delta_b$, we must take $NH(\delta_b)$ rounds of parity measurement in $Z$
basis. Note that the number of likely phase string, $s_p$ on those remained
untagged
pairs is now not fixed, since it is determined by the phase string 
of all pairs ($S_p$) 
instead of the phase string untagged pairs only ($s_p$). 
More specifically, at any round of parity measurement
in $Z-$basis, if the destination pair is untagged, the the number of
likely string $s_p$ is unchanged. But if the destination pair is tagged,
the number of likely string $s_p$ for the remained untagged pairs is doubled:
the destination pair has half a chance to bear a phase-flip error. Actually,
the assumption of half a phase-flip error of the tagged qubit is also 
the worst-case assumption: this maximizes
the number of likely phase string $s_p$ for the remained untagged pairs.   
After the bit-flip error corrections are completed, the number of
likely string $s_p$ for the remained untagged pairs is
\begin{eqnarray}
\omega_{p,untagged}= 2^{(1-\Delta^2)N\cdot H(\delta_p/(1-\Delta))}.
\end{eqnarray}  
Here we have used the fact that among all destination pairs, a fraction 
$\Delta$ of them had been tagged.
\\Now we can consider how to do the privacy amplification (phase-flip correction). 
Straightly, we 
assume half a phase-flip for each tagged pairs and can just use the increased
phase-flip error rate ($\delta_p+\Delta/2$) 
and complete the distillation. But our purpose here is only to do privacy 
amplification rather than entanglement purification. 
Using the method giving by ref\cite{gllp}, we 
can treat the issue more sophisticatedly.
Since we only want to obtain the final key, it makes no difference for 
Alice and Bob to measure each remained pairs in $Z-$basis
 before privacy amplification.
Suppose they have done so. 
We now see what happens after the privacy amplification
is activated. In each step of privacy amplification, they randomly choose
a subset $Sub_j$. They then randomly choose an amplifying bit $d_j$ in 
set $Sub_j$. They replace each bit value $v_k$ in $Sub_j$ by 
$v_k\oplus v_D$ and $v_D$ is the bit value of $d_j$. They discard
bit $d_j$. They repeat such operations for $l$ times. If they are 
sure that among all $l$ amplifying (discarded) bits, at least 
$\log_2 \omega_{p,untagged}$ of them are originally from untagged pairs,
then those remained bits originally from untagged pairs are perfectly
secure. Since this means a separate privacy amplification has been
taken to the untagged bits. Explicitly, after $l$ rounds,
we denote the set of the 
remained bits originally 
from untagged pairs by $\{b_{ui}\}$, we have
\begin{eqnarray}
b_{ui}=v_{ui}\oplus v_{ti} 
\end{eqnarray}   
and $v_{ui}$ is the resultant value due to independent 
privacy amplification which $only$ happens to all untagged bits, 
$v_{ti}$ is
the parity of certain tagged bits. Since  $v_{ui}$ itself is perfectly secure,
$b_{ui}$ is also secure. Therefore, after $l$ rounds of privacy amplification,
among all of the remained bits, $1-\Delta$ of them are unconditionally secure while
$\Delta$ of them could be still insecure. The fraction $\Delta$ are those 
bits which
are originally from tagged pairs. In particular,
\begin{eqnarray}
l=\frac{\log_2 \omega_{p,untagged}}{1-\Delta}=(1+\Delta)N\cdot H\left(\frac{\delta_p}{1-\Delta}\right).
\end{eqnarray}
And there are
\begin{eqnarray}
q=N\left[1-H(\delta_b)-H\left(\frac{\delta_p}{1-\Delta}\right)-
\Delta H\left(\frac{\delta_p}{1-\Delta}\right)\right]
\end{eqnarray}
bits remained. The next task is to remove those $\Delta q$ insecure bits. 
To do so they can simply
continue the same privacy amplification for another $\Delta q$ rounds. 
After additional
$\Delta q$ rounds are taken, each remained bit has the form of
$b_{ui}'\oplus v_{ti}'$ and $\{b_{ui}'\}$ and each elements in $\{b_{ui}'\}$ are just the parity 
of certain subset of $b_{ui}$ and all $b_{ui}'$ are independent. 
The final key rate is
\begin{eqnarray}
\nonumber R_f =1-\Delta  -(1-\Delta)H(\delta_b)\\- H\left(\frac{\delta_p}{1-\Delta}\right) 
+\Delta^2 H\left(\frac{\delta_p}{1-\Delta}\right). 
\end{eqnarray}
Our protocol directly applies to the source of weak coherent states. 
In our protocol, Alice may choose to measure all of her qubits in 
the begining and tells
Eve the outcome of a fraction of them. Given the source of 
weak coherent states, since the phase of each signal is random, it is just an
imperfect single-photon source that produces multi-photon signals 
occasionally. Here Alice does not tell Eve any outcome, but the multi-photon
signals play the role of tagged qubits, given the PNS attack.
  
{\it Summary} In summary, we have given a clear picture on how to do
error correction and privation amplification with two $independent$
parity matrices and all  computational
difficulties in practical QKD are removed. 
We have also applied our method to the case of QKD with imperfect
source and given a formula for key rate.
\acknowledgments
I am
very grateful to Prof. H. Imai for his long term supports.

\end{document}